\documentclass[useAMS,usenatbib]{mn2e}

\def \aap{A\&A}

\def \apjl{ApJ}
\def \apjs{ApJS}

\usepackage{color}
\usepackage{graphicx}
\usepackage{mathabx}
\usepackage[T1]{fontenc}
\usepackage{aecompl}
\usepackage{txfonts}

\title[High Order Harmonics in Kepler Light Curves]{High Order Harmonics in Light Curves of Kepler Planets}
\author[Caden Armstrong and Hanno Rein ]{Caden Armstrong$^{1,2}$\thanks{caden.armstrong@mail.utoronto.ca} and Hanno Rein$^{1}$\\
$^{1}$University of Toronto, Department of Environmental and Physical Sciences, 1265 Military Trail, Toronto Ontario M1C 1A4, Canada\\
$^{2}$Centre for Planetary Sciences, University of Toronto, 1265 Military Trail, Toronto Ontario M1C 1A4, Canada.}
\begin{document}

\date{Draft version: \today{}}

\pagerange{\pageref{firstpage}--\pageref{lastpage}} \pubyear{2015}

\maketitle
\label{firstpage}

\begin{abstract}
The Kepler mission was launched in 2009 and has discovered thousands of planet candidates.
In a recent paper, \cite{esteves} found a periodic signal in the light curves of KOI-13 and HAT-P-7, with a frequency triple the orbital frequency of a transiting planet.
We found similar harmonics in many systems with a high occurrence rate.
At this time, the origins of the signal are not entirely certain.

We look carefully at the possibility of errors being introduced through our data processing routines but conclude that the signal is real.
The harmonics on multiples of the orbital frequency are a result of non-sinusoidal periodic signals. 
We speculate on their origin and generally caution that these harmonics could lead to wrong estimates of planet albedos, beaming mass estimates, and ellipsoidal variations.
\end{abstract}


\section{Introduction}\label{sec:introduction}
To date, the Kepler mission has discovered over one thousand confirmed planets and many more planet candidates\footnote{http://kepler.nasa.gov/}. 
The main goal of the mission was to discover Earth-like planets but many other scientific results have made use of this high quality dataset \citep{kepler}.
For example, \cite{elip} used the Kepler dataset to measure ellipsoidal variations which can be used to derive masses and radii of both planets and stars. 

In a recent paper, \cite{esteves} found a mysterious third harmonic in the phase curves of Kepler-13b and HAT-P-7b.
Until now, this signal had been detected in few planetary systems.
Previously, tidal effects were the only known source of this effect \citep{morris}.
In our own investigation, we observe these higher order harmonics in a number of systems, including Kepler-13b, but tidal effects do not provide a sufficient explanation. 
The mostly likely source of these signals, is due to non-sinusoidal periodic variations in the light curves of these planets.

The outline of the paper is as follows.
First, we describe the methods and results in Sect.~\ref{sec:methods}. 
In Sect.~\ref{sec:error} we discuss and dismiss the possibility of introducing this signal in our pipeline. 
The analysis of tidal effects are covered in Sect.~\ref{sec:tidal}
We summarize and conclude our results in Sect.~\ref{sec:conclusions}.

\section{Method and Results}\label{sec:methods}
\begin{figure*}
\includegraphics[width=\textwidth]{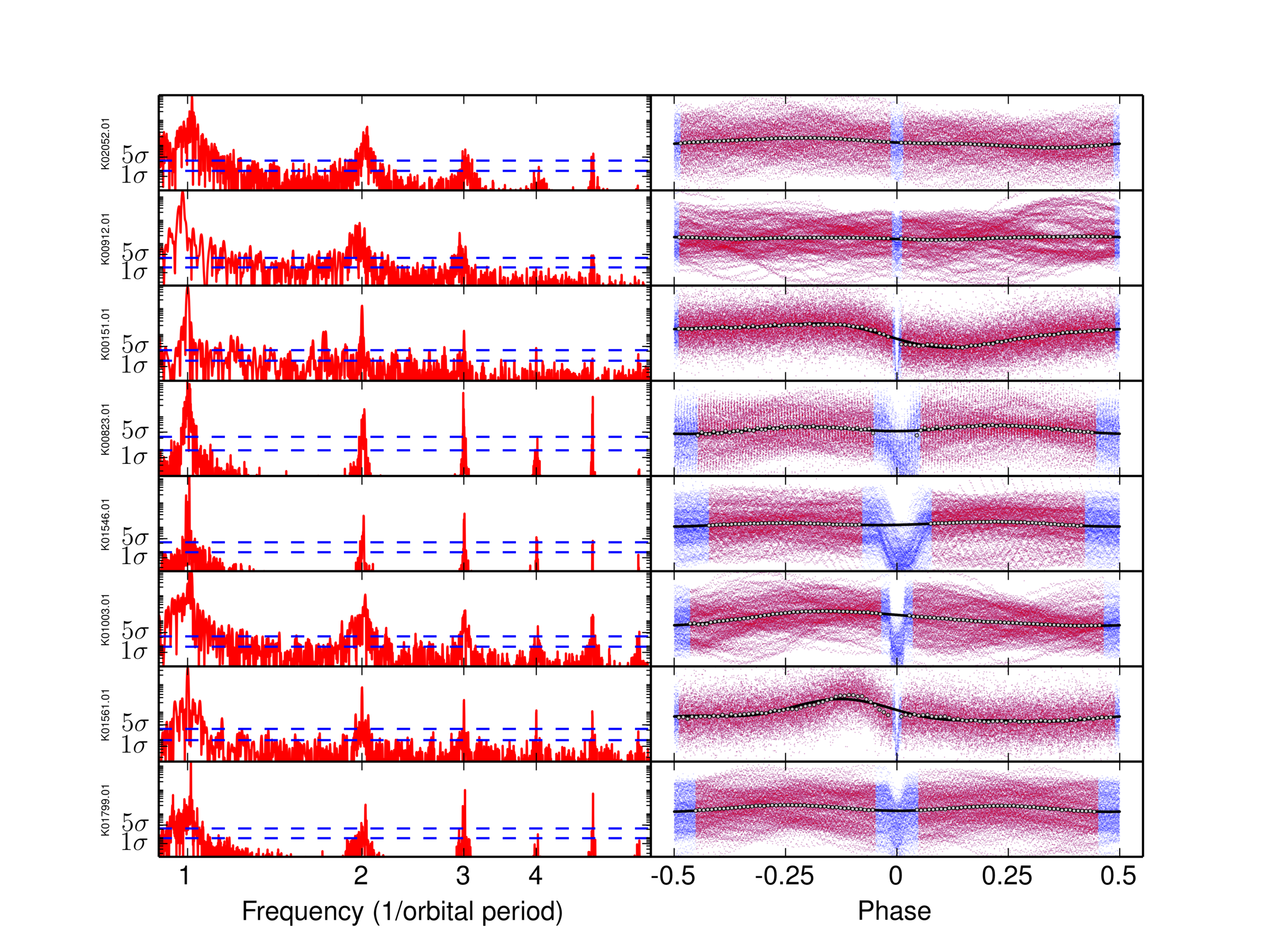}
\caption{The left panel show the periodograms plotted in a log scale, with power in y-axis and the false alarm probabilities plotted as the dotted lines. The figures on the right are the folded light curves for illustrative purposes, showing the removed sections of data. The plot scaling is set to only show analyzed points, both the transits and outlying data points may extend outside the visible area of the plot. Blue data points were removed prior to creating the power spectra. The black line is the fitted three harmonics model, and the grey dots are binned data points. \label{fig:periodigrama}}
\end{figure*}
\begin{figure*}
\includegraphics[width=\textwidth]{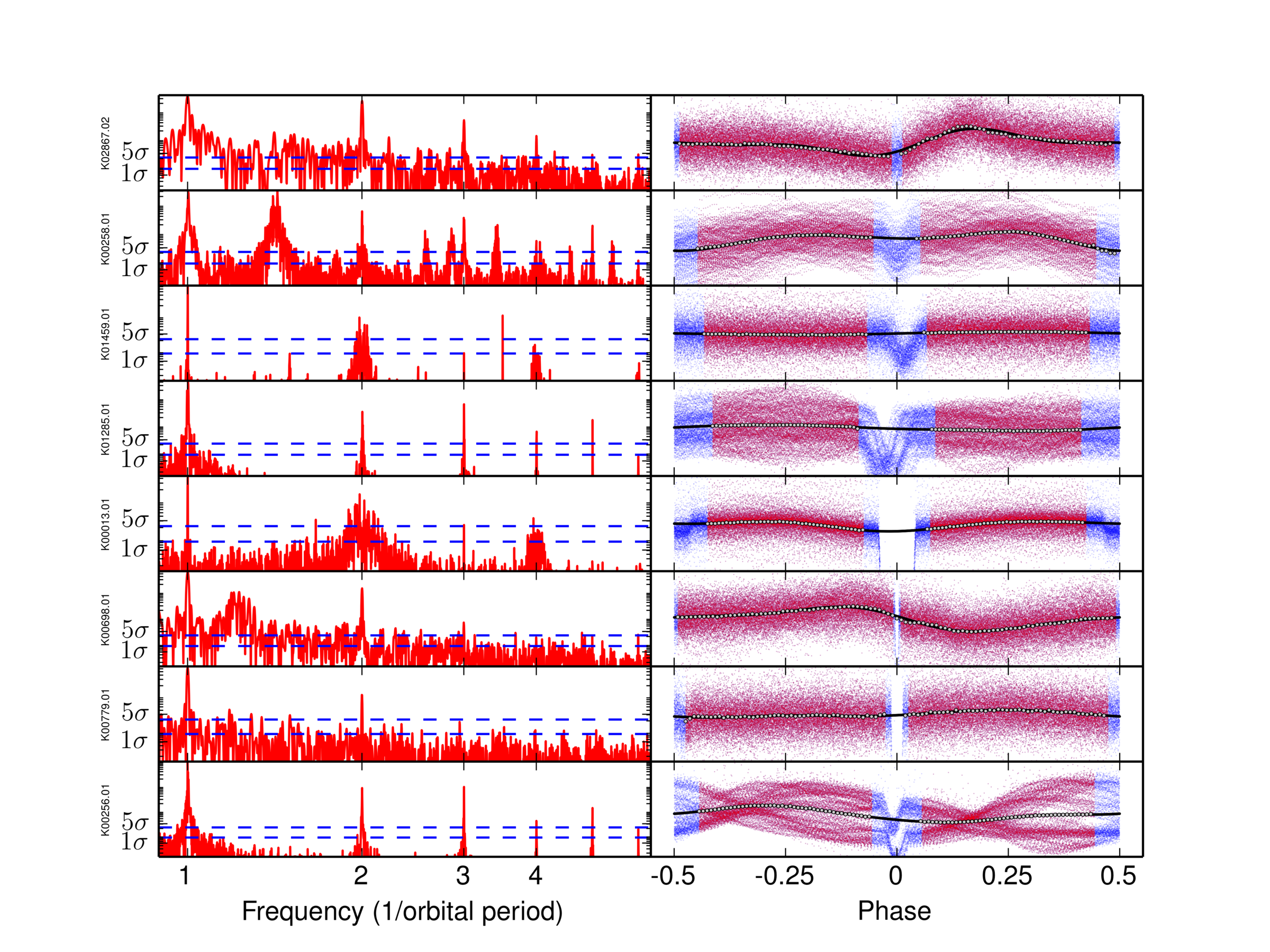}
\caption{Same as Fig.~\ref{fig:periodigrama} \label{fig:periodigramb}}
\end{figure*}

We worked with publicly available data of Kepler light curves \citep{kepler}.
The planets selected were those with the \textit{confirmed} tag in the Kepler Object of Interest (KOI) table from the Mikulski Archive for Space Telescopes (MAST)\footnote{Updated March 1, 2013}.
We looked at the long cadence data set including quarters~0-17.
The data was retrieved from the \textit{.fits} files using the PyFits module, available in the AstroPy project \citep{astropy}.

The Kepler team provides different data products for each system.
Here, we investigate the \texttt{PDCSAP\_FLUX} and \texttt{SAP\_FLUX} columns.
The \texttt{SAP\_FLUX} column contains the light curve data produced using simple aperture photometry which was produced by the Photometric Analysis module of the NASA Science Operations Center pipeline. 
The column provides flux in electrons per second contained in the optimal aperture pixels, collected by the Kepler spacecraft.
The \texttt{PDCSAP\_FLUX} column contains the Photometric Analysis data after the Pre-Search Data Conditioning (PDC) algorithm has been applied.
The PDC module applies a series of corrections to the data for known instrumental anomalies, along with unanticipated artifacts\footnote{Kepler Data Characteristics Handbook. 31 May 2013.}.
Our analysis was run on both the \texttt{PDCSAP\_FLUX} and \texttt{SAP\_FLUX} columns with little to no effect on our final results.

The flux for each planet was normalized to the average flux in each quarter.
Normalizing to the average provided the simplest and most consistent method of combining the different quarters without introducing artificial signals.
The values for planet period, transit midpoint, and transit width were taken from the MAST Kepler catalogue.
The separate light curves from each quarter were then merged together to create a single light curve.

All data points falling within the primary and secondary transit of each planet were removed from the light curves.
The width of the transits was increased by a factor of two to account for any transit timing variations, duration variations, or small errors in the cutting process. 
By comparison, the transit duration variation as measured in \cite{Mazeh} are only upwards of a few percent.
Note that cutting out all transits with a longer transit duration reduces the signal to noise ratio but proved to be the simplest method of ensuring that the transits would not contribute to the signal (see also discussion in Sect.~\ref{sec:error}).
The folded light curves are shown on the right panel of Fig.~\ref{fig:periodigrama} and Fig.~\ref{fig:periodigramb}.
Data points in blue are those near the primary or secondary transit and have been cut.
In addition to the transits, outlying data points were removed.

To verify the period of the signal, we use a power spectrum.
We turn to a Lomb-Scargle \citep{lomb} periodogram as the data points are unevenly spaced.
Lomb-Scargle periodograms were created using the module available in the SciPy package \citep{scipy}.
The periodograms were sampled in an evenly spaced log scale from $10^{-0.05}$ to $10^{0.8}$ orbital frequencies, with 2000 sample points.
The power spectra were plotted, and analyzed by eye, looking for any spikes at frequencies that correspond to multiples of the orbital frequency.
Only spectra that had a high peak compared to the overall noise were considered.
This left us with 16 systems with clearly identifiable signals with a high signal to noise ratio.
Their properties are listed in Table~\ref{tab:planetdata}.
The signal found by \cite{esteves} in HAT-P-7 was not recovered.

The periodograms for these systems are plotted in the left panels of Fig.~\ref{fig:periodigrama} and Fig.~\ref{fig:periodigramb}.
They show the following general features: 
\begin{enumerate}
\item A peak on the orbital frequency of the planet, attributed to Doppler boosting \citep[also known as beaming,][]{beaming} and planetary reflection \citep{beer}.
\item A second peak on twice the orbital frequency, attributed to the ellipsoidal deformation of the star under the effects of the planet \citep{elip}.
\item Additional harmonics at multiples of the orbital frequency of unknown origins.
\end{enumerate}
Of the 16 systems, the third harmonic is easily visible in 13 of them.
In all of these systems, the periodograms showed narrow peaks exactly at the triple orbital frequency.
In a few of these systems, similar peaks are found at higher orbital frequencies.
The width of the peaks for these harmonics are of similar size compared to the known lower order harmonics.
It is important to note that many of the systems showing the signal have been mentioned in the two papers that look at transit timing variations of Kepler planets.
The following planets are mentioned in \cite{Mazeh}: 
KOI-823.01, 
KOI-1546.01,
KOI-1459.01,
and KOI-1285.01.
The following are mentioned in \cite{szabo}:
KOI-256.01,
KOI-823.01.
While not all of the transit timing variations were measurable in \cite{Mazeh}, they were still concluded to be significant.
It is also interesting to note that in \cite{Mazeh}, the third harmonic for KOI-203.01 is visible in their Fig.~17 but not discussed any further. 
In our own analysis, the signal to noise ratio was not significant enough in KOI-203.01 for signals to be visible.

\begin{table*}
\begin{tabular}{|ccccccc|}
KOI Name & Period [days] & Depth [${\rm ppm}$] & Planet radius [$r_\Earth$] & Multiplicity & TTV Period [days] & TTV Amp [min]\\
\hline\hline
K00013.01 &$1.76359 \pm 0.000001$&$4633$&$23 \pm 16.0$&1 &	 $   $ & $   $ \\ \hline
K00151.01 &$13.4472 \pm 0.000047$&$1426$&$6 \pm 2.2$&1 &	 $   $ & $   $ \\ \hline
K00256.01 &$1.37868 \pm 0.000002$&$26406$&$24 \pm 12.0^*$&1 &	 $47.5^B$ & $0.44^B$ \\ \hline
K00258.01 &$4.15745 \pm 0.000006$&$1136$&$5.1 \pm 2.3$&1 &	 $   $ & $   $ \\ \hline
K00698.01 &$12.7187 \pm 0.000006$&$7514$&$10.87 \pm 0.29$&1 &	 $   $ & $   $ \\ \hline
K00779.01 &$10.406 \pm 0.000007$&$14610$&$12.1 \pm 5.1$&1 &	 $   $ & $   $ \\ \hline
K00823.01 &$1.02841 \pm 0.000004$&$5867$&$8.7 \pm 3.8$&1 &	 $184^A$ & $0.85^A$ \\ \hline
K00912.01 &$10.8485 \pm 0.000031$&$1651$&$2.3 \pm 0.29$&2 &	 $   $ & $   $ \\ \hline
K01003.01 &$8.36058 \pm 0.000005$&$19878$&$10.8 \pm 6.1$&1 &	 $   $ & $   $ \\ \hline
K01285.01 &$0.937444 \pm 0.000003$&$4971$&$7 \pm 2.3$&1 &	 $-^{AC}$ & $-^{AC}$ \\ \hline
K01459.01 &$0.692007 \pm 0.000001$&$3876$&$4.12 \pm 0.29$&1 &	 $-^{AC}$ & $-^{AC}$ \\ \hline
K01546.01 &$0.917547 \pm 0.000002$&$14150$&$9.5 \pm 2.9$&1 &	 $-^{AC}$ & $-^{AC}$ \\ \hline
K01561.01 &$9.08594 \pm 0.00004$&$2184$&$26 \pm 16.0$&1 &	 $   $ & $   $ \\ \hline
K01799.01 &$1.73108 \pm 0.000002$&$12542$&$49 \pm 31.0$&1 &	 $   $ & $   $ \\ \hline
K02052.01 &$4.22464 \pm 0.000028$&$1026$&$4.1 \pm 2.0$&1 &	 $   $ & $   $ \\ \hline
K02867.02 &$18.9407 \pm 0.0014$&$14$&$0.36 \pm 0.15$&2 &	 $   $ & $   $ \\ \hline
\end{tabular}
\caption{ Planet parameters.  Values used from the Kepler Objects of Interest table from MAST.
${}^A$\protect\cite{Mazeh}.${}^B$\protect\cite{szabo}.
${}^C$ No value was given in the respective paper, but transit timing variations were still found to be significant.
\label{tab:planetdata} }
\end{table*}

In the MAST table for the Kepler Objects of Interest, the chosen planets were all of the \texttt{CONFIRMED} status, as of March 1st, 2013.
In \cite{KOI256}, the Kepler Object of Interest KOI-256 was concluded not to be a planetary system, but a mutually eclipsing binary consisting of a cool white dwarf and an active M3 Dwarf.
The expected transit is in fact the occultation of the white dwarf passing behind the M dwarf.
The period for the white dwarf is unchanged from the value reported in the MAST table.
In their analysis, higher order harmonics of the white dwarf are found in the light curve of the system.

\section{Artifact or real signal}\label{sec:error}
The first possibility of the signal's origins resides in the misinterpretation of the periodograms.
It is easy to assume that the signal is produced as a combination of the known signals (beaming, ellipsoidal variation, etc.). 
However, there is no way to combine the lower order periodic signals on the period of 1 orbital period and $\frac{1}{2}$ orbital period, to create the higher order harmonic.
To check for these types of errors computationally, synthetic light curves were generated with the known signals, and analyzed with the same pipeline used to process the Kepler data.
From these tests, no higher order harmonics could be recovered.

There is the obvious possibility that we introduced artificial signals in the light curves by cutting out transits.
Any kind of Fourier analysis such as the Lomb-Scargle periodogram is susceptible to aliasing.
To investigate this issue, we applied our transit cutting methods twice, once at the actual orbital frequency and once at a frequency that is an irrational multiple of the orbital frequency, $\nu_{\rm injected} = 1.3575656\; \nu_{\rm orbit}$.
This value was chosen such that neither $\nu_{\rm injected}$ nor any of its harmonics fall near the orbital frequency or its harmonic, thus obscuring any artificial signals that we were attempting to create.
The light curves were then prepared (normalized and cut) in the same fashion as for the original analysis.
From this point, the periodograms were created with the same methods described previously, and were examined for newly created signals.
In the power spectra, we were unable to detect artificial signals created by our cutting method near the frequency $\nu_{\rm injected}$ or any multiple thereof.
Thus, we can rule out our data preparation methods for the origins of the higher order harmonic.

We investigated one more possibility that could lead to artifacts in the periodograms.
A potential issue might arise for systems of high eccentricity. 
In such systems, secondary transits can be shifted significantly from where we would expect them for a more circular orbit.
This would cause the transit to bleed over and not be cut properly.
With these extra data-points now contributing to the periodogram, higher order signals may be introduced.
Similarly, transit timing variation could cause the same issue.
We ran several tests with larger transit windows.
We ranged our transit durations from a factor of $1.5$ to a factor of $3$ of the transit width.
In all of these tests, the higher order harmonics did not disappear, but expectedly the signal to noise ratio was lower.

Not all signatures in the light curves may necessarily be sinusoidal.
We tested generating synthetic light curve data containing both sinusoidal and non-sinusoidal periodic signals.
From this test, higher order harmonics can be recovered with varying results.
As expected, the shape of the non-sinusoidal signal determines the relative powers of the higher order harmonics.
Due to the nature of these signals, amplitude measurements for ellipsoidal variation, and beaming can be significantly affected.
These effects are important as they have recently been used to determine masses of planets.
We note that in \cite{shporer}, beaming measurements were found not to agree with radial velocity measurements for mass estimates.
\section{Tidal Effects}\label{sec:tidal}
If the third harmonic was entirely tidal effects, it would be expected to be similar to the effect on a single star in a binary system as described by \cite{morris}:
\begin{equation}
a_{3c} = -25\mu_\star (2+\tau_\star)(\frac{R_\star}{A})^4 (\frac{M_p}{M_\star}) \sin^3 i / 32(3-\mu_\star)
\end{equation}
Where $a_{3c}$ is the cosine term of the harmonic on triple the frequency of the planet. 
$\mu_\star$ is the linear limb darkening coefficient, $A$ is the semi-major axis of the planet, $i$ is the inclination of the planet's orbit, $M_p$ and $M_\star$ are the masses for the planet and star, $\tau_\star$ is the gravity darkening coefficient, and $R_\star$ is the radius of the star.
Using a least-squares fit with a three-harmonic model, we measured the amplitude of the third harmonic.
In \cite{morris} tidal effects between binary stars were expected to produce high order harmonics at frequencies that are multiples of the orbital period.
These tidal effects were applied to Kepler-13b in \cite{shporer}, to describe the high order harmonics found.
The third order signal was found to have twice the amplitude expected in the \cite{morris} model.
For the systems analyzed in this paper, all overshoot the expected amplitude.
This brings up two possibilities: tidal effects between a planet-star and star-star are different enough to warrant further analysis, or there is an additional unknown astrophysical effect.
It is unlikely to be entirely tidal, as many of the signals are many orders of magnitude larger than expected, and uncorrelated with the value expected from theory.
\begin{table}
\begin{tabular}{|ccc|}
KOI Name & Expected $a_{3c}$ & Fitted $a_{3c}$\\
\hline\hline
K00013.01$^A$ & $ 2.61 \cdot 10^{-7} $ & $2.39\cdot 10^{-6} $\\ \hline 
K00151.01 & $ 1.33 \cdot 10^{-11} $ & $5.95\cdot 10^{-6} $\\ \hline 
K00256.01$^B$ & - & $6.00\cdot10^{-3}$               \\ \hline  
K00258.01 & $ 1.03 \cdot 10^{-9} $ & $8.90\cdot 10^{-5} $\\ \hline 
K00698.01 & $ 3.24 \cdot 10^{-11} $ & $7.20\cdot 10^{-6} $\\ \hline 
K00779.01 & $ 1.90 \cdot 10^{-9} $ & $2.65\cdot 10^{-6} $\\ \hline 
K00823.01 & $ 8.03 \cdot 10^{-9} $ & $9.88\cdot 10^{-6} $\\ \hline 
K00912.01 & $ 2.55 \cdot 10^{-11} $ & $3.24\cdot 10^{-5} $\\ \hline 
K01003.01 & $ 3.27 \cdot 10^{-9} $ & $5.10\cdot 10^{-3}$  \\ \hline
K01285.01 & $ 1.16 \cdot 10^{-8} $ & $2.06\cdot 10^{-5}$\\ \hline 
K01459.01 & $ 2.44 \cdot 10^{-8} $ & $1.47\cdot 10^{-5} $\\ \hline 
K01546.01 & $ 4.88 \cdot 10^{-7} $ & $3.00\cdot 10^{-4}$                \\ \hline
K01561.01 & $ 1.97 \cdot 10^{-8} $ & $7.52\cdot 10^{-5} $\\ \hline 
K01799.01 & $ 1.04 \cdot 10^{-5} $ & $5.95\cdot 10^{-3}$              \\ \hline
K02052.01 & $ 7.92 \cdot 10^{-15} $ & $8.10\cdot 10^{-3}$                \\ \hline
K02867.02 & $ 2.93 \cdot 10^{-19} $ & $2.01\cdot 10^{-5} $\\ \hline 
\end{tabular}
\caption{ The amplitudes for both the expected and fitted term of the cosine term for the third harmonic are in the form of relative flux.
The expected amplitudes were calculated with the parameters from the Kepler Catalogue.
The mass of many of these planets is unknown, so to approximate the mass ratios, the radii were used with the assuming that the star and planet would have similar density.
All of the errors in the expected values are of the same order as the value.
${}^A$ The value quoted is for consistency with the other planets. 
If the mass ratio from \protect\cite{shporer} is used, a similar factor of 2 is found.
${}^B$ The KOI-256 system contains a white dwarf, for which the limb darkening coefficients are unknown and assumed to be zero.  \label{tab:tidalchart} }
\end{table}

\section{Conclusions}\label{sec:conclusions}
We analyzed the light curves of 16 Kepler systems.
Through the use of simple Lomb-Scargle periodigrams, we were able to find evidence for the beaming effect, ellipsoidal variations and reflection, as previously extracted successfully by other groups.
In addition, of the 16 chosen planets, 13 exhibited higher order harmonics in the light curve, on frequencies that are multiples of the orbital frequency for which there is currently no explanation.
Through careful testing, we conclude that the higher order harmonics found do not originate from our own data processing.
Thus, we are left with the following possible explanations:

{\bf Tidal effects.}
By applying the analysis in \cite{morris}, we conclude that tidal effects cannot be the sole contributor to high order harmonics.
The measured values of the high order harmonics exceeds what tidal effects predict by a range of a factor of 2 up to many orders of magnitude.
It is possible that tidal effects between planet-star and star-star are significantly different, but unlikely to make up for the huge orders of magnitude.

{\bf Non-sinusoidal periodic light curve variations.}
In our tests, we found that non-sinusoidal variations can indeed replicate large amplitude higher harmonics, and depending on the shape of the variations, different combinations of higher order harmonics can be excited.
Non-sinusoidal light curves might for example be caused by non-isotropic reflection or thermal emission from the planet's surface.

{\bf An error in the Kepler pipeline}
Although unlikely, there could in principle be an error introduced early on in the Kepler pipeline. 
A more sophisticated pixel-level analysis of Kepler data could provide further insight into the signal's origin.

At the present, we speculate that the most likely cause of these high order harmonics are non-sinusoidal periodic light curve variations.
We note that there is a weak correlation between the amplitude of the high order harmonics and the planet's radius, but no obvious correlation to period.
If the physical origin of these harmonics is fully understood, we are likely to gain new insight into the planetary properties.
Until then however, one should exercise caution when determining planetary properties from the shape of the light curve as the harmonics of unknown origin affect the modelling of known astrophysical phenomena.

\section*{Acknowledgements}
{\small 
Caden Armstrong thanks the Centre for Planetary Sciences and the University of Toronto for a summer undergraduate research fellowship.
Hanno Rein and Caden Armstrong acknowledge support from the NSERC Discovery grant RGPIN-2014-04553.
This research has made use of the NASA Exoplanet Archive, which is operated by the California Institute of Technology, under contract with the National Aeronautics and Space Administration under the Exoplanet Exploration Program.
This paper includes data collected by the Kepler mission. Funding for the Kepler mission is provided by the NASA Science Mission directorate.
This research made use of Astropy, a community-developed core Python package for Astronomy \citep{astropy}.
}
\bibliography{harmonicbib.bib}
\appendix

\end{document}